\documentclass[11pt,a4paper]{article}
\usepackage[T1]{fontenc}
\usepackage[utf8]{inputenc}
\usepackage{lmodern}
\usepackage[margin=25mm]{geometry}
\usepackage{amsmath,amssymb}
\usepackage{bm}
\usepackage{graphicx}
\usepackage{verbatim}
\usepackage{booktabs}
\usepackage[numbers,sort&compress]{natbib}
\usepackage{microtype}
\usepackage{xurl}
\usepackage[hidelinks]{hyperref}
\title{Prediction intervals for overdispersed multinomial data with application to historical controls}
\author{%
  S\"oren Budig$^{1}$\thanks{Corresponding author: Department of Biostatistics,
  Institute of Cell Biology and Biophysics, Leibniz University Hannover.
  Email: \href{mailto:budig@cell.uni-hannover.de}{\texttt{budig@cell.uni-hannover.de}}}
  \and Frank Schaarschmidt$^{1}$
  \and Max Menssen$^{2,3}$}
\date{}
\hypersetup{
  pdftitle={Prediction intervals for overdispersed multinomial data with application to historical controls},
  pdfauthor={S\"oren Budig, Frank Schaarschmidt, Max Menssen}
}

\begin{document}
\maketitle
\begin{center}
\small
\begin{minipage}{0.96\textwidth}
\centering
$^{1}$ Department of Biostatistics, Institute of Cell Biology and Biophysics,
Leibniz University Hannover\par\smallskip
$^{2}$ Department of Medical Statistics,
University Medical Center G\"ottingen (UMG)\par\smallskip
$^{3}$ Lower Saxony Center for Artificial Intelligence and Causal Methods
in Medicine (CAIMed), Hannover, Germany
\end{minipage}
\end{center}

\begin{abstract}
In pharmaceutical and toxicological research, historical control data is increasingly being used to validate concurrent control groups, typically via the construction of historical control limits. Although methods have been described for continuous, dichotomous and count data, approaches for overdispersed multinomial endpoints, common in developmental and reproductive toxicology or histopathology, are currently lacking. We describe and compare frequentist and Bayesian methods for constructing simultaneous prediction intervals for one future multinomial count vector. Frequentist methods include normal approximations and parametric bootstrap procedures with symmetric, asymmetric, marginal, or rank-based calibration. Bayesian methods use hierarchical Dirichlet-multinomial models with two alternative hyperpriors. A simulation study varies the number and size of historical clusters, category probabilities and overdispersion, and we assess simultaneous coverage and marginal lower- and upper-tail error probabilities. Normal-approximation methods frequently produce undercoverage, particularly with few historical studies or small expected counts. Marginal bootstrap calibration and rank-based intervals generally achieve coverage closer to the nominal level and more balanced marginal tail errors. The performance of Bayesian methods depends on the hyperprior, with conservatism being particularly evident near the multinomial case. An application to historical rat lung histopathology data illustrates prediction of the severity distribution among affected animals. These methods fill an important gap for multinomial endpoints and support the validation of concurrent controls using historical control data, in line with the recent European Food Safety Authority scientific opinion on the use and reporting of historical control data.
\end{abstract}

\section{Introduction}

Multinomial data, representing outcomes classified into more than two mutually exclusive, unordered categories, arise frequently in different research fields, including clinical trials, epidemiology, genomics, and toxicology. For example, in developmental and reproductive toxicology (DART) studies the effects of pharmaceuticals on the growth and development of an organism, and on the reproductive system are assessed \cite{NTP2020DART01}. Developmental toxicity is assessed, for example, by examining dose-dependent effects on fetal development. After administration of different doses of a substance, pregnant mice are sacrificed after a certain period of time and the offspring are classified into categories such as `alive', `malformed' or `dead', yielding multinomial data. Typically, such studies use the same experimental design, with an untreated control group and several treatment groups, and a specific model organism, for example a particular rat strain. The experimental protocols are, in general, defined by guidelines \cite{OPPTS1998, EMA2020}. Consequently, as standardized toxicological studies are repeated, the amount of the historical control information increases.

In regulatory toxicology and preclinical research, there is growing interest in utilizing historical control data (HCD) from previous, similar studies \cite{Greim2003, Elmore2009, Hayashi2011, Dertinger2023, Menssen2023, EFSA2025}. HCD can serve various purposes \cite{Kluxen2021}, including: quality assurance of the test system \cite{OED2014}, identification of unreasonable concurrent controls \cite{Weber2011}, assessment of the biological relevance of observed effects \cite{KluxenJensen2021}, and informally addressing the multiple comparison problem \cite{Kluxen2020}. Furthermore, incorporating HCD can increase statistical power \cite{Bonapersona2021, Gurjanov2023}, contextualize results (especially for rare events like specific tumors or malformations), and reduce the sample size required for concurrent controls \cite{Schmidli2014, Steger2020, Gurjanov2024b, Menssen2026}, aligning with ethical considerations and the 3Rs principles (Reduction, Refinement, Replacement). However, the use of HCD introduces challenges, primarily the risk of bias due to systematic differences between historical and current studies arising from factors like changes in animal strains (genetic drift), husbandry practices, diagnostic criteria, or laboratory environments over time \cite{Hatswell2020}. 

While HCD can fulfill multiple functions, this manuscript focuses on validating the concurrent control group. This is typically achieved by calculating historical control limits to determine whether current observations fall within the central $100(1-\alpha)\%$ of the underlying historical distribution or are off-trend \cite{Menssen2023}. Ad hoc methods, such as the historical range \cite{Lovell2020}, empirical quantiles \cite{Dertinger2023,Kluxen2021} or the mean value $\pm$ 2 standard deviations \cite{Levy2019}, are often used, though inadequate for this purpose \cite{Menssen2025}. Instead, the use of prediction intervals (PIs), which should cover the future observation with probability $1-\alpha$, is recommended \cite{Menssen2019, Kluxen2021, Menssen2022, EFSA2025}

Prediction intervals based on HCD have been developed for several data types \cite{Francq2019, Menssen2022, Menssen2024}, including overdispersed binomial endpoints \cite{Menssen2019,Menssen2025}. Here, we extend and compare methods for constructing intervals for overdispersed multinomial endpoints, with particular emphasis on simultaneous historical control limits and marginal tail balance.

Two main challenges arise in the multinomial setting. First, prediction intervals must be constructed for multiple categories simultaneously, so multiplicity must be accounted for to ensure an overall simultaneous coverage probability of $1-\alpha$. Second, these data are often overdispersed, meaning that the variability in the data is greater than expected under the standard multinomial distribution \cite{Hinde1998, Hilbe2011}. In the context of HCD, data are hierarchically structured, with observations nested within study groups. Between-study variability, caused by differences in personnel, environment, or animal batches, can induce positive correlation among observations within the same control group \cite{Palazzi2024, Zarn2024}, necessitating statistical methods that account for this extra-multinomial variation.

In this article, we present and compare methods for constructing PIs for overdispersed multinomial data. We first describe methodologies for modeling such data, followed by procedures for constructing PIs. We evaluate the performance of these methods via a simulation study regarding coverage probability and interval balance, and finally, apply them to a real-world toxicological dataset.

\section{Methods}

\subsection{Data structure}

We consider $K$ independent historical studies (clusters), indexed by $k=1,\ldots,K$, with a common mean category-probability vector. Each cluster $k$ has a fixed sample size $n_k$. Experimental units within a cluster are assigned to one of $C$ mutually exclusive categories ($c=1,\ldots,C$). This leads to a historical count column vector $\bm{x}_{k}=(x_{k1},x_{k2},\ldots,x_{kC})^T$, whose components satisfy $\sum_{c=1}^{C}x_{kc}=n_k$. As a baseline without overdispersion, we consider the multinomial model: 
\begin{equation}
    \bm{x}_{k}\sim \textnormal{Multinomial}(n_{k},\bm{\pi}),
\end{equation}
where $\bm{\pi} = (\pi_1, \pi_2, \ldots, \pi_C)^T$ is a vector consisting of the probabilities that an experimental unit falls into category $c$. The mean and covariance are:
\[
    \textnormal{E}(\bm{x}_{k})   =n_{k}\bm{\pi},
\]
\[
    \textnormal{Cov}(\bm{x}_{k}) = {\bm{\Sigma}}_{k},   
\]
where ${\bm{\Sigma}}_{k}$ denotes the multinomial variance-covariance matrix, expressed as ${\bm{\Sigma}}_{k}=n_{k}\{\textnormal{Diag}(\bm{\pi})-\bm{\pi}\bm{\pi}^T\}$ \cite{McCullaghNelder1989}.

\subsection{Overdispersed multinomial data}\label{sec:overdispersed_multinomial_data}

Experimental units within the same historical study are often positively correlated, leading to overdispersion. Several approaches exist to account for overdispersion in multinomial data, including extended distributions like the Dirichlet-multinomial (DM) \cite{Mosimann1962, Morel1993, Zhang2016, CorsiniViroli2022}, mixed models with random effects \cite{Hartzel2001, Hedeker2003, Chan2023}, and Generalized Estimating Equations \cite{Touloumis2013}. Here, we employ a quasi-likelihood approach \cite{McCullaghNelder1989}, which retains $\textnormal{E}(\bm{x}_k)=n_k\bm{\pi}$ and scales the standard multinomial covariance matrix by a common dispersion parameter $\phi\geq 1$:
\begin{equation}
    \textnormal{Cov}(\bm{x}_{k}) =\phi {\bm{\Sigma}}_{k}.
\end{equation}
Various estimators for $\phi$ have been proposed for sparse multinomial data \cite{Farrington1996, Fletcher2012, DengPaul2016, AfrozFletcher2020}. We utilize the estimator $\hat{\phi}_A$ proposed by Afroz, Parry, and Fletcher~\cite{AfrozFletcher2020}, which has demonstrated strict error control in the multiple comparison settings investigated by Budig, Vogel, and Schaarschmidt~\cite{Budig2026}. It is defined as:
\begin{equation}
    \hat{\phi}_A = \frac{\chi^2 / \text{df}}{1+\overline{s}},
\end{equation}
where $\chi^2$ is the Pearson statistic:
\[
    \chi^2=\sum_{k=1}^{K}\sum_{c=1}^{C}\frac{(x_{kc}-n_{k}\hat{\pi}_{c}){}^2}{n_{k}\hat{\pi}_{c}}.
\]
The residual degrees of freedom are $\text{df}=KC-K-P$, where $P$ denotes the number of non-redundant parameters \cite{McCullaghNelder1989}. If no other covariates or factors are present in the model, $P$ can be computed using $P=C-1$.

The term $\overline{s}$ is a bias correction, calculated as:
\[
    \overline{s}=\sum_{k=1}^{K}\sum_{c=1}^{C}((x_{kc}-n_{k}\hat{\pi}_{c})/n_{k}\hat{\pi}_{c})/(KC-K).
\]

\subsection{Data generation using the Dirichlet-Multinomial Distribution}\label{sec:data_generation}

While we adopt the quasi-likelihood approach (Section~\ref{sec:overdispersed_multinomial_data}) for parameter estimation, motivated by the multiple-comparison results of Budig, Vogel, and Schaarschmidt~\cite{Budig2026}, we employ the Dirichlet-Multinomial (DM) distribution to generate synthetic data with controlled overdispersion. This generation process is utilized for both the parametric bootstrap procedures (Section~\ref{sec:prediction_interval_methods}) and the simulation study (Section~\ref{sec:simulation_study}).

The DM distribution extends the multinomial distribution \cite{Mosimann1962} by assuming that the probability vector $\bm{\pi}_k$ for each cluster $k$ is a random variable drawn from a DM with parameter vector $\bm{\eta} = (\eta_1, \dots, \eta_C)^T$, where $\eta_c > 0$. Formally:
\begin{align}
    \bm{\pi}_k &\sim \textnormal{Dirichlet}(\bm{\eta}) \\
    \bm{x}_k | \bm{\pi}_k  &\sim \textnormal{Multinomial}(n_k, \bm{\pi}_k)
\end{align}
The expected count is $\textnormal{E}(\bm{x}_k) = n_k \textnormal{E}(\bm{\pi}_k) = n_k \frac{\bm{\eta}}{\eta_0}$, where $\eta_0 = \sum_{c=1}^{C} \eta_c$. The covariance structure is given by:
\begin{equation}
    \textnormal{Cov}(\bm{x}_k) = n_k \left\{ \textnormal{Diag}(\bm{\pi}) - \bm{\pi}\bm{\pi}^T \right\} \left( \frac{n_k + \eta_0}{1 + \eta_0} \right),
\end{equation}
where $\bm{\pi} = \textnormal{E}(\bm{\pi}_k)$ \cite{Johnson1997}. The term $\phi_{\textnormal{DM}} = \frac{n_k + \eta_0}{1 + \eta_0}$ represents the overdispersion factor. This formulation explicitly incorporates overdispersion, as $\phi_{\textnormal{DM}} > 1$ for finite $\eta_0>0$ and $n_k>1$. As $\eta_0 \to \infty$, $\phi_{\textnormal{DM},k} \to 1$ and the DM distribution converges to the multinomial distribution.

To simulate data with a target probability vector $\bm{\pi}_{\text{true}}$, a specific overdispersion $\phi$, and equal cluster sizes ($n_k=n$), the parameter $\eta_0$ is derived by rearranging the expression for the common factor $\phi_{\textnormal{DM}}$:
\begin{equation}
    \eta_0 = \frac{n - \phi}{\phi - 1}.
\end{equation}
The individual Dirichlet parameters are then set as $\bm{\eta} = \eta_0 \bm{\pi}_{\text{true}}$. For a finite positive $\eta_0$, the required range is $1<\phi<n$. Notably, with uniform cluster sizes, data generated via this process satisfy the quasi-multinomial assumption of a constant dispersion factor $\phi$.

The data generating process for each cluster $k$ is as follows. First, a cluster-specific probability vector $\bm{\pi}_k$ is sampled from a Dirichlet distribution, defined as $\bm{\pi}_k \sim \textnormal{Dirichlet}(\eta_0 \bm{\pi}_{\text{true}})$. Subsequently, the observation vector $\bm{x}_k$ is drawn from a multinomial distribution conditioned on this probability vector, such that $\bm{x}_k \sim \textnormal{Multinomial}(n_k, \bm{\pi}_k)$.

\subsection{Prediction interval methods}\label{sec:prediction_interval_methods}

This section details the statistical methods employed for constructing simultaneous prediction intervals (PIs) for future multinomial observations based on historical control data (HCD). Because historical control limits are required for all $C$ categories, there arises the problem of multiplicity. The intervals must jointly achieve simultaneous coverage rather than only category-wise coverage. We aim to construct simultaneous PIs  $[L_c, U_c]$ for categories $c=1,\dots,C$ such that a vector of future observations $\bm{y}=(y_1,\dots,y_C)^T$ with fixed sample size $m=\sum_{c=1}^{C}y_c$ is contained within these limits with probability approximately $1-\alpha$, or conservatively at least that level::
\[
P(L_c \le y_c \le U_c, \; \forall c) \approx 1 - \alpha.
\]
In addition to simultaneous coverage, it is desirable for two-sided PIs to be balanced, meaning that the probability of prediction error is distributed equally between the lower and upper tails, yielding equal tail probabilities. Ideally, for any category $c$, the marginal probability of falling below the lower limit ($P(y_c < L_c)$) and exceeding the upper limit ($P(y_c > U_c)$) should be approximately equal. While standard methods often construct symmetrical intervals, multinomial data with small counts or probabilities near 0 or 1 often exhibit skewness. Consequently, we also consider methods that calibrate the lower and upper bounds separately to achieve equal tail probabilities.

A preliminary step for all methods involves fitting a intercept-only multinomial model to the HCD to obtain estimates of the underlying probabilities $\hat{\bm{\pi}}=(\hat{\pi}_1,\dots,\hat{\pi}_C)^T$ and the overdispersion parameter $\hat{\phi}$. The expected count for each category in a future sample of size $m$ is estimated as $\hat{y}_c=m\hat{\pi}_c$. 

However, a potential future observation $y_c$ will deviate from this expected value due to random sampling variability and uncertainty in the estimated parameters. This deviation is defined as the prediction error:
\[
    y_c - \hat{y}_c.
\]
To construct intervals that achieve the nominal simultaneous coverage probability, we must quantify the variability of this prediction error distribution. The prediction standard error for each category $c$, denoted $\textnormal{sep}_c$, captures this variability by combining the variance of a random future observation $y_c$ and the variance of the estimator $\hat{y}_c$. Because $\hat{y}_c = m\hat{\pi}_c$, we obtain $\textnormal{Var}(\hat{y}_c)=m^2\textnormal{Var}(\hat{\pi}_c)$. For independent historical clusters with common dispersion $\phi$, the pooled estimator has variance $\phi\pi_c(1-\pi_c)/N_{\textnormal{hist}}$, where $N_{\textnormal{hist}}$ denotes the total historical sample size. This yields the following expression, analogous to the binomial case described by Menssen and Schaarschmidt~\cite{Menssen2019}:
\begin{align}
    \widehat{\textnormal{Var}}(y_c) &= \hat{\phi}m\hat{\pi}_c(1-\hat{\pi}_c), \nonumber \\
    \widehat{\textnormal{Var}}(\hat{y}_c) &= \frac{\hat{\phi}m^2\hat{\pi}_c(1-\hat{\pi}_c)}{N_{\textnormal{hist}}}, \nonumber \\
    \textnormal{sep}_c &= \sqrt{\widehat{\textnormal{Var}}(y_c)+\widehat{\textnormal{Var}}(\hat{y}_c)}.
\end{align}
The frequentist methods described in Sections~\ref{sec:pointwise}--\ref{sec:scsrank} construct the PIs $[L_c, U_c]$ by scaling the prediction standard error $\textnormal{sep}_c$ with suitable quantiles or calibrated multipliers.

\subsubsection{Pointwise Normal Approximation}\label{sec:pointwise}

Following the general normal-approximation approach to PIs described by Nelson~\cite{Nelson1982}, we approximate the standardized prediction error for each category independently by a standard normal distribution. The interval for category $c$ is:
\[
    [L_c, U_c]=[\hat{y}_c - z_{1-\alpha/2}\cdot \textnormal{sep}_c,\quad \hat{y}_c + z_{1-\alpha/2}\cdot \textnormal{sep}_c],
\]
where $z_{1-\alpha/2}$ is the standard normal quantile. However, these intervals do not guarantee simultaneous coverage for all C categories, and subsequent methods address this primary goal more directly.

\subsubsection{Bonferroni Adjustment}

To address the lack of simultaneous coverage in the pointwise approach, the simplest solution is the Bonferroni adjustment, which constructs each individual interval at a stricter confidence level. The error rate for each category is adjusted to $\alpha/C$. The intervals are calculated as:
\[
    [L_c, U_c]=[\hat{y}_c - z_{1-\alpha/(2C)} \cdot \textnormal{sep}_c,\hat{y}_c + z_{1-\alpha/(2C)} \cdot \textnormal{sep}_c].
\]
Theoretically, the Bonferroni inequality ensures that the simultaneous coverage probability is at least $1-\alpha$. However, this lower bound holds only if the marginal intervals have correct coverage, which depends on the validity of the normal approximation here.

\subsubsection{Multivariate Normal Approximation}

To improve upon the pointwise approach, we can account for the inherent correlation between the counts of the $C$ categories by approximating the prediction error vector with a multivariate normal (MVN) distribution. Using the covariance rules for random vectors \cite{Hardle2024}, the covariance matrix of the prediction error vector $\bm{y}-\hat{\bm{y}}$, denoted $\bm{\Sigma}_{\textnormal{pred}}$, is given by
\[
    \textnormal{Cov}(\bm{y}-\hat{\bm{y}})=\textnormal{Cov}(\bm{y})+\textnormal{Cov}(\hat{\bm{y}})-\textnormal{Cov}(\bm{y},\hat{\bm{y}})-\textnormal{Cov}(\hat{\bm{y}},\bm{y}).
\]
Because the future observation $\bm{y}$ is assumed independent of the estimator $\hat{\bm{y}}$ obtained from the historical data, the two cross-covariance terms are zero. Hence,
\[
    \bm{\Sigma}_{\textnormal{pred}} = \textnormal{Cov}(\bm{y}) + \textnormal{Cov}(\hat{\bm{y}}).
\]
Under the quasi-multinomial model, we have $\textnormal{Cov}(\bm{y}) = \phi m (\textnormal{Diag}(\bm{\pi}) - \bm{\pi}\bm{\pi}^T)$ and $\textnormal{Cov}(\hat{\bm{y}}) = \frac{\phi m^2}{N_{\textnormal{hist}}}(\textnormal{Diag}(\bm{\pi}) - \bm{\pi}\bm{\pi}^T)$. Substituting the parameter estimates yields:
\[
    \widehat{\bm{\Sigma}}_{\textnormal{pred}} = \widehat{\textnormal{Cov}}(\bm{y}) + \widehat{\textnormal{Cov}}(\hat{\bm{y}}) = \hat{\phi} m \left(1 + \frac{m}{N_{\textnormal{hist}}}\right) (\textnormal{Diag}(\hat{\bm{\pi}}) - \hat{\bm{\pi}}\hat{\bm{\pi}}^T).
\]
The simultaneous critical value $q^{\textnormal{mvn}}$ is obtained as the $(1-\alpha)$ equi-coordinate two-sided quantile of a $C$-variate standard normal distribution with correlation matrix $\widehat{\bm{R}}_{\textnormal{pred}}$, which is derived by standardizing $\widehat{\bm{\Sigma}}_{\textnormal{pred}}$ \cite{mvtnorm}. The two-sided simultaneous PI for the $c$-th category is:
\[
    [L_c, U_c]=[\hat{y}_c - q^{\textnormal{mvn}} \cdot \textnormal{sep}_c, \quad \hat{y}_c + q^{\textnormal{mvn}} \cdot \textnormal{sep}_c ].
\]

\subsubsection{Symmetric Bootstrap Calibration}\label{sec:sym_calib}

Bootstrap calibration offers a flexible alternative to asymptotic approximations. This method replaces the theoretical quantile with an empirically calibrated multiplier, $q^{\textnormal{sym}}$, that is chosen to achieve the nominal simultaneous coverage probability. The algorithm was adapted from Menssen and Rathjens~\cite{Menssen2025} for multinomial data. Because it uses one common symmetric multiplier across all categories, the method is simple and directly targets simultaneous coverage, but it does not adapt to category-specific skewness and may therefore produce less balanced intervals when the prediction error distributions are asymmetric.

\begin{enumerate}
    \item Use the historical data $\bm{x}_k$ to estimate the parameters $\hat{\bm{\pi}}$ and $\hat{\phi}$. For DM data generation, the dispersion parameter must satisfy $1 < \phi < s$, where $s$ denotes the sample size of the draw being generated. Therefore, if $\hat{\phi} \le 1$, set $\hat{\phi}=1.01$. If $\hat{\phi}$ is greater than or equal to the relevant sample size, truncate it to a value slightly below that size. For truncation, we use $0.975s$ for a draw of size $s$.
    \item Generate $B$ bootstrap replicates. For each $b=1, \dots, B$:
    \begin{itemize}
        \item Generate a historical dataset $\bm{X}^*_b = \{\bm{x}^*_{1,b}, \dots, \bm{x}^*_{K,b}\}$ from a DM distribution using the historical cluster sizes $n_k$ and parameters $\hat{\bm{\pi}}, \hat{\phi}$.
        \item If a category $c$ in $\bm{X}^*_b$ contains zero counts across all clusters (i.e., $\sum_k x^*_{k,c,b} = 0$), add a single count to a randomly selected cluster for that category to ensure model convergence.
        \item Generate a future observation vector $\bm{y}^*_b$ from a DM distribution using the future sample size $m$ and parameters $\hat{\bm{\pi}}, \hat{\phi}$.
    \end{itemize}
    \item For each bootstrap dataset $\bm{X}^*_b$, re-estimate the parameters to obtain $\hat{\bm{\pi}}^*_{b}$ and $\hat{\phi}^*_{b}$.
    \item Calculate the bootstrapped expected future counts $\hat{y}^*_{b,c} = m\hat{\pi}^*_{b,c}$ and the bootstrapped prediction standard errors $\textnormal{sep}^*_{b,c}$ for each category $c$ within replicate $b$
    \item Use a bisection algorithm to find the calibrated multiplier $q^{\textnormal{sym}}$. This algorithm iteratively adjusts a candidate multiplier until the empirical simultaneous coverage probability lies within a pre-specified tolerance $t$ of the nominal level $1-\alpha$. Specifically, determine $q^{\textnormal{sym}}$ such that:
    \[
        \left| \frac{1}{B} \sum_{b=1}^{B} \mathbb{I} \left( \hat{y}_{b,c}^{*} - q^{\textnormal{sym}} \cdot \textnormal{sep}_{b,c}^{*} \le y_{b,c}^{*} \le \hat{y}_{b,c}^{*} + q^{\textnormal{sym}} \cdot \textnormal{sep}_{b,c}^{*} \quad \forall \quad c \right) - (1-\alpha) \right| \le t,
    \]
    where $\mathbb{I}(\cdot)$ is the indicator function.
\end{enumerate}

The final symmetrical simultaneous PIs are constructed using the calibrated multiplier:
\[
    [L_{c},U_{c}] = \left[ \hat{y}_{c} - q^{\textnormal{sym}} \cdot \textnormal{sep}_c, \quad \hat{y}_{c} + q^{\textnormal{sym}} \cdot \textnormal{sep}_c \right].
\]

\subsubsection{Asymmetric Bootstrap Calibration}

To account for potential skewness in the prediction error distribution, particularly when cluster sizes are small and the category-wise errors are skewed in a similar direction, this method calibrates separate multipliers for the lower ($q^{\textnormal{asy}}_{L}$) and upper ($q^{\textnormal{asy}}_{U}$) bounds. This allows the interval width to vary asymmetrically around the predicted value. However, because the method uses one common lower and one common upper multiplier for all categories, it is less suitable when the multinomial vector contains both rare categories ($\pi_c \approx 0$) and a dominant category ($\pi_c \approx 1$), since the corresponding prediction errors may be skewed in opposite directions.

The procedure utilizes the same bootstrap samples generated in the symmetrical approach:

\begin{enumerate}
    \item Perform Steps 1--4 of the bootstrap procedure described in Section~\ref{sec:sym_calib}.
    
    \item Use two separate bisection algorithms to find the calibrated multipliers $q^{\textnormal{asy}}_{L}$ and $q^{\textnormal{asy}}_{U}$. The algorithms adjust these multipliers until the simultaneous coverage probability for each bound lies within a tolerance $t$ of the target $1-\alpha/2$.
    \begin{itemize}
        \item Find $q^{\textnormal{asy}}_{L}$ such that:
        \[
            \left| \frac{1}{B} \sum_{b=1}^{B} \mathbb{I} \left( \hat{y}_{b,c}^{*} - q^{\textnormal{asy}}_{L} \cdot \textnormal{sep}_{b,c}^{*} \le y_{b,c}^{*} \quad \forall \quad c \right) - (1-\alpha/2) \right| \le t.
        \]
        \item Find $q^{\textnormal{asy}}_{U}$ such that:
        \[
            \left| \frac{1}{B} \sum_{b=1}^{B} \mathbb{I} \left( y_{b,c}^{*} \le \hat{y}_{b,c}^{*} + q^{\textnormal{asy}}_{U} \cdot \textnormal{sep}_{b,c}^{*} \quad \forall \quad c \right) - (1-\alpha/2) \right| \le t.
        \]
    \end{itemize}
    
\end{enumerate}

The final asymmetrical simultaneous PIs are constructed as:
\[
    [L_{c}, U_{c}] = \left[ \hat{y}_{c} - q^{\textnormal{asy}}_{L} \cdot \textnormal{sep}_c, \quad \hat{y}_{c} + q^{\textnormal{asy}}_{U} \cdot \textnormal{sep}_c \right].
\]

\subsubsection{Marginal Bootstrap Calibration}

This approach calibrates category-specific multipliers for the lower ($q^{\textnormal{marg}}_{c,L}$) and upper ($q^{\textnormal{marg}}_{c,U}$) limits. Unlike the previous methods, which use global multipliers, this method calibrates each lower and upper limit separately for each category to a Bonferroni-adjusted level. As a result, it can accommodate category-specific skewness in either direction and is therefore well suited to multinomial settings with heterogeneous probability patterns, including combinations of rare categories and dominant categories.

The procedure utilizes the same bootstrap samples generated in the symmetrical approach:

\begin{enumerate}
    \item Perform Steps 1--4 of the bootstrap procedure described in Section~\ref{sec:sym_calib}.
    
    \item For each category $c=1,\dots,C$, use separate bisection algorithms to find the calibrated multipliers $q^{\textnormal{marg}}_{c,L}$ and $q^{\textnormal{marg}}_{c,U}$. The algorithms adjust these multipliers until the marginal coverage probability for each bound lies within a tolerance $t$ of the target $1-\alpha/(2C)$.
    \begin{itemize}
        \item Find $q^{\textnormal{marg}}_{c,L}$ such that:
        \[
            \left| \frac{1}{B} \sum_{b=1}^{B} \mathbb{I} \left( \hat{y}_{b,c}^{*} - q^{\textnormal{marg}}_{c,L} \cdot \textnormal{sep}_{b,c}^{*} \le y_{b,c}^{*} \right) - \left(1-\frac{\alpha}{2C}\right) \right| \le t.
        \]
        \item Find $q^{\textnormal{marg}}_{c,U}$ such that:
        \[
            \left| \frac{1}{B} \sum_{b=1}^{B} \mathbb{I} \left( y_{b,c}^{*} \le \hat{y}_{b,c}^{*} + q^{\textnormal{marg}}_{c,U} \cdot \textnormal{sep}_{b,c}^{*} \right) - \left(1-\frac{\alpha}{2C}\right) \right| \le t.
        \]
    \end{itemize}
\end{enumerate}

The PIs for each category $c$ are constructed using these individually calibrated multipliers:
\[
    [L_c, U_c]=[\hat{y}_c - q^{\textnormal{marg}}_{c,L} \cdot \textnormal{sep}_c,\quad \hat{y}_c + q^{\textnormal{marg}}_{c,U} \cdot \textnormal{sep}_c].
\]

\subsubsection{Maximum Absolute Studentized Residual}\label{sec:max_abs_studentized}

This method is an adaptation of the studentized bootstrap \cite{Efron1993}. Specifically, a single simultaneous critical value is obtained from the empirical distribution of the maximum absolute studentized residual across all categories. By studentizing the residuals, the method places categories with different prediction variances on a common scale and directly targets simultaneous coverage. However, because it uses one common symmetric cutoff to all categories, it does not adapt to category-specific skewness and may therefore yield less balanced intervals when prediction error distributions differ substantially across categories.

The procedure utilizes the same bootstrap samples generated in the previous approaches:
\begin{enumerate}
    \item Perform Steps 1--4 of the bootstrap procedure described in Section~\ref{sec:sym_calib}.
    \item For each bootstrap replicate $b$ and each category $c$, a pivotal quantity is calculated as the studentized residual:
    \begin{equation*}
    z_{b,c} = \frac{y_{b,c}^* - \hat{y}_{b,c}^*}{\textnormal{sep}_{b,c}^*}
    \end{equation*}
    \item For each bootstrap replicate $b$, determine the maximum absolute deviation across all categories:
    \[
        z_{b, \textnormal{max}} = \max_{c=1,\dots,C} |z_{b,c}|.
    \]
    This yields a vector of $B$ maximum statistics $\bm{z}_{\textnormal{max}} = (z_{1, \textnormal{max}}, \dots, z_{B, \textnormal{max}})$.
    \item Determine the simultaneous critical value, $q^{\textnormal{masr}}$, as the $(1-\alpha)$-quantile of the empirical distribution of $\bm{z}_{\textnormal{max}}$.
\end{enumerate}

The resulting simultaneous PIs for each category $c$ is constructed by scaling the original prediction standard error by this single critical value:
\[
    [L_c, U_c]=\left[ \hat{y}_c - q^{\textnormal{masr}} \cdot \textnormal{sep}_c, \quad \hat{y}_c + q^{\textnormal{masr}} \cdot \textnormal{sep}_c \right].
\]

\subsubsection{Rank-Based SCS}\label{sec:scsrank}

We adapt the rank-based rectangular simultaneous-set construction of Besag et al.~\cite{Besag1995} to prediction residuals. We retain the label Rank-Based SCS for this adaptation, although its target is a future observation rather than a model parameter. Like the previous bootstrap approaches, it is based on studentized residuals, but it derives the critical values from their ranks rather than from their absolute magnitudes. As a result, the method can adapt to category-specific error distributions without imposing symmetry and is therefore well suited to multinomial settings with heterogeneous skewness across categories.

The procedure utilizes the same bootstrap samples generated in the previous approaches:
\begin{enumerate}
    \item Follow Step 1 and 2 of Section~\ref{sec:max_abs_studentized} to compute the $B \times C$ matrix $\bm{Z}$ of studentized residuals, where the entry for replicate $b$ and category $c$ is $z_{b,c}$.
    \item For each category $c$ (i.e., column-wise), rank the residuals from smallest to largest. Let $r_{b,c}$ denote the rank of $z_{b,c}$ among the $B$ bootstrap replicates for category $c$.
    
    \item For each bootstrap replicate $b$ (i.e., row-wise), calculate an extremeness score $w_b$:
    \[
        w_b = \max \left( \max_{c=1,\dots,C}(r_{b,c}), \quad B + 1 - \min_{c=1,\dots,C}(r_{b,c}) \right).
    \]
    
    \item Determine the critical rank $\tau^*$ as the $(1-\alpha)$-quantile of the extremeness scores $w_b$. Specifically, if $w_{(1)} \le \dots \le w_{(B)}$ are the ordered scores, then $\tau^* = w_{(k)}$, where $k$ is the integer nearest to $(1-\alpha)B$.
    
    \item Map the critical rank back to the specific residuals for each category. Let $z_{c,(1)} \le \dots \le z_{c,(B)}$ denote the ordered residuals for category $c$. The lower and upper critical values for category $c$ are:
    \begin{align*}
        q^{\textnormal{rank}}_{c,L} &= -z_{c, (B + 1 - \tau^*)}, \\
        q^{\textnormal{rank}}_{c,U} &= z_{c, (\tau^*)}.
    \end{align*}
\end{enumerate}    

The final simultaneous PIs are constructed by scaling the original prediction standard errors with these category-specific critical values:
\[
    [L_c, U_c]=\left[ \hat{y}_c - q^{\textnormal{rank}}_{c,L} \cdot \textnormal{sep}_c, \quad \hat{y}_c + q^{\textnormal{rank}}_{c,U} \cdot \textnormal{sep}_c \right].
\]

\subsubsection{Bayesian Prediction Intervals}

As an alternative to the frequentist framework, PIs can be constructed using a Bayesian hierarchical model. This approach directly models the structure of the historical data, including the between-study heterogeneity, and generates intervals based on the posterior predictive distribution of a future observation.

We assume the historical count vectors $\bm{x}_k$ for studies $k=1, \dots, K$ arise from a DM process. The model is specified as follows:
\begin{align*}
    \bm{\pi}_k | \bm{\pi}_{\textnormal{global}}, \eta_0 &\sim \textnormal{Dirichlet}(\eta_0 \cdot \bm{\pi}_{\textnormal{global}}), \\
    \bm{x}_k | \bm{\pi}_k &\sim \textnormal{Multinomial}(n_k, \bm{\pi}_k).
\end{align*}
Here, $\bm{\pi}_{\textnormal{global}}$ is the population mean probability vector and $\eta_0>0$ is a concentration parameter. Historical and future clusters are assumed independent conditional on these hyperparameters. The corresponding intraclass correlation is $\rho=1/(1+\eta_0)$ \cite{Landsman2019}, and a cluster of size $s$ has dispersion $1+(s-1)\rho$.

We assign independent priors to the population mean probability vector and the heterogeneity parameter. For the mean vector, we use
$\bm{\pi}_{\textnormal{global}}\sim\textnormal{Dirichlet}(\bm{1}_C)$, which is uniform on the probability simplex \cite{Frigyik2010}. To explore sensitivity to the heterogeneity prior, we compare two illustrative specifications. First, motivated by the use of half-Cauchy priors for hierarchical scale parameters \cite{Polson2012}, we assign $\eta_0\sim\textnormal{Cauchy}^+(0,5)$. Its heavy upper tail permits large concentrations corresponding to weak overdispersion, while retaining support for small concentrations associated with substantial between-study heterogeneity. Second, we assign $\rho\sim\textnormal{Beta}(1,10)$, which favors smaller within-study correlations. Specifying a prior directly on the intraclass correlation facilitates interpretation of assumptions about within-study dependence \cite{Spiegelhalter2001}, and Beta priors have also been suggested for this parameter in beta-binomial models \cite{JansenHolling2023}.

The model is fitted to the complete historical dataset $\bm{X} = \{\bm{x}_1, \dots, \bm{x}_K\}$ using Markov Chain Monte Carlo (MCMC) simulation in Stan \cite{cmdstan} (4 chains, each with 1,250 warm-up iterations followed by 2,500 retained sampling iterations) to obtain samples from the joint posterior $p(\bm{\pi}_{\textnormal{global}}, \eta_0 | \bm{X})$. From these samples, we generate the posterior predictive distribution for a future observation $\bm{y}$ of size $m$. For each posterior draw $s=1,\dots,S$, we first draw a latent probability vector for the new study, $\tilde{\bm{\pi}}^{(s)} \sim \textnormal{Dirichlet}(\eta_0^{(s)} \cdot \bm{\pi}_{\textnormal{global}}^{(s)})$, and then draw the future observation $\bm{y}^{(s)}_{\textnormal{pred}} \sim \textnormal{Multinomial}(m, \tilde{\bm{\pi}}^{(s)})$. This results in a collection of $S$ predicted vectors $\{\bm{y}^{(s)}_{\textnormal{pred}}\}$ representing the predictive distribution $p(\bm{y} | \bm{X})$.

We utilize these predictive samples to construct simultaneous PIs using three distinct approaches, all of which were included in the subsequent simulation study:

\begin{enumerate}
    \item The first approach defines the interval limits using the marginal quantiles of the posterior predictive distribution. To ensure simultaneous coverage, we apply a Bonferroni correction. For each category $c$, the lower and upper bounds are set to the $\alpha/(2C)$ and $1 - \alpha/(2C)$ quantiles, respectively, of the sample set $\{y^{(1)}_{\textnormal{pred},c}, \dots, y^{(S)}_{\textnormal{pred},c}\}$.
    
    \item The second approach constructs symmetric simultaneous intervals centered on the posterior predictive mean $\hat{\bm{y}}_{\textnormal{bayes}}$ (the column-wise mean of the samples). Analogous to the frequentist Maximum Absolute Studentized Residual method (Section~\ref{sec:max_abs_studentized}), we scale the intervals using a single critical value derived from the distribution of maximum standardized deviations. We first compute the posterior predictive standard deviation, $\textnormal{sd}_{\textnormal{bayes}, c}$, for each category. Then, for each posterior sample $s$, we calculate the maximum standardized deviation across all categories:
    \[
        z_{s, \textnormal{max}} = \max_{c=1,\dots,C} \left| \frac{y^{(s)}_{\textnormal{pred},c} - \hat{y}_{\textnormal{bayes},c}}{\textnormal{sd}_{\textnormal{bayes}, c}} \right|.
    \]
    The critical value $q^{\textnormal{bayes}}$ is defined as the $(1-\alpha)$-quantile of the vector $\bm{z}_{\textnormal{max}} = (z_{1, \textnormal{max}}, \dots, z_{S, \textnormal{max}})$. The intervals are given by $[L_c, U_c]=\left[ \hat{y}_{\textnormal{bayes},c} \pm q^{\textnormal{bayes}} \cdot \textnormal{sd}_{\textnormal{bayes}, c} \right]$.

    \item The third approach adapts the rank-based procedure from Section~\ref{sec:scsrank} to the Bayesian context. We treat the $S$ posterior predictive samples as analogues to bootstrap replicates and rank the posterior predictive counts themselves, category by category, across these samples. The resulting extremeness scores are then used to select lower and upper bounds directly from the posterior predictive count distribution, yielding rectangular simultaneous posterior predictive intervals. Thus, unlike the frequentist rank-based procedure, the Bayesian interval is constructed from the raw posterior predictive counts rather than from studentized statistics.
    
\end{enumerate}

\subsection{Simulation Study}\label{sec:simulation_study}

We conducted a simulation study to evaluate the performance of the proposed bootstrap and Bayesian methods alongside standard frequentist approximations for constructing simultaneousp PIs for overdispersed multinomial data. The primary objective was to assess the empirical simultaneous coverage probability under a range of conditions pertinent to toxicological studies.

We systematically varied several factors to assess their impact on the performance of the PI methods. The number of historical clusters was set to $K \in \{5,10,20,100\}$ to reflect scenarios with both limited and extensive historical databases. The historical cluster sizes ($n_k = n$) and the future sample size ($m$) were assumed equal, with $n=m \in \{10, 50, 100, 500\}$. To investigate performance across varying degrees of overdispersion, three levels of dispersion parameter were simulated: $\phi \in \{1.01,5,8\}$, where 1.01 represents a near-multinomial case. Finally, 32 distinct true probability vectors ($\bm{\pi}_{\text{true}}$) were considered, comprising 12 scenarios for $C=3$, 10 for $C=5$ and 10 for $C=10$. These vectors ranged from balanced settings with equal category probabilities to highly imbalanced rare-event scenarios, with individual category probabilities ranging from 0.01 to 0.98. Complete tables of all probability vectors are given in Appendix~\ref{sec:AppMPSS}.

For each parameter combination, 1,000 simulation iterations were performed. In each iteration, historical control data $\bm{x}_k$ and a corresponding future observation $\bm{y}$ were generated from a DM distribution with cluster size $n$, dispersion $\phi$, and probability vector $\bm{\pi}_{\text{true}}$ (see Section~\ref{sec:data_generation}). All methods were applied to this same adjusted historical dataset. The independent future vector remained unmodified. We then constructed PIs with nominal coverage 95\%. For the bootstrap-based approaches, $B=10,000$ parametric bootstrap samples were drawn per simulation run. The Bayesian models were fitted in Stan using 4 chains with 2,500 sampling iterations each, as described in Section~\ref{sec:prediction_interval_methods}.

The primary performance metric was the empirical simultaneous coverage probability, defined as the proportion of simulation runs in which the future observation vector $\bm{y}$ was simultaneously contained within the PIs for all categories (i.e., $P(L_c \le y_c \le U_c, \; \forall c=1, \dots, C)$). A method was considered to perform well if its empirical coverage probability approximated the nominal level of $1-\alpha=0.95$.

In addition to overall simultaneous coverage, we assessed the interval balance by calculating the marginal tail error probabilities for each category $c=1, \dots, C$. Specifically, we evaluated the empirical marginal probability of under-prediction, $P(y_c < L_c)$, and over-prediction, $P(y_c > U_c)$. For a method with well-controlled, symmetric error rates, these marginal tail probabilities should be approximately equal for a given category. 

The simulations were performed using the R software version 4.4.3 \cite{Rcitation}. Multinomial models were fitted using the \texttt{VGAM} package \cite{Yee2015}. The MVN approximation was implemented via the \texttt{mvtnorm} package \cite{mvtnorm}, and Dirichlet random variates were generated using \texttt{MCMCpack} \cite{MCMCpack}. Parallelization was achieved using the \texttt{future.apply} package \cite{future.apply}. The simulation code and the code used to evaluate the subsequent histopathological example are available at \url{https://github.com/sbudig/predint_multnom}.

\subsection*{Use of artificial intelligence tools}

During the preparation of this manuscript, the authors used OpenAI's ChatGPT (primarily GPT-6.1 Sol) to assist with proofreading and improvements to grammar, spelling, and linguistic clarity of the text. All suggested changes were reviewed by the authors, who take full responsibility for the final manuscript.

\section{Results}

\begin{figure}[!tbp]
    \centering
    \includegraphics[width=\linewidth,keepaspectratio]{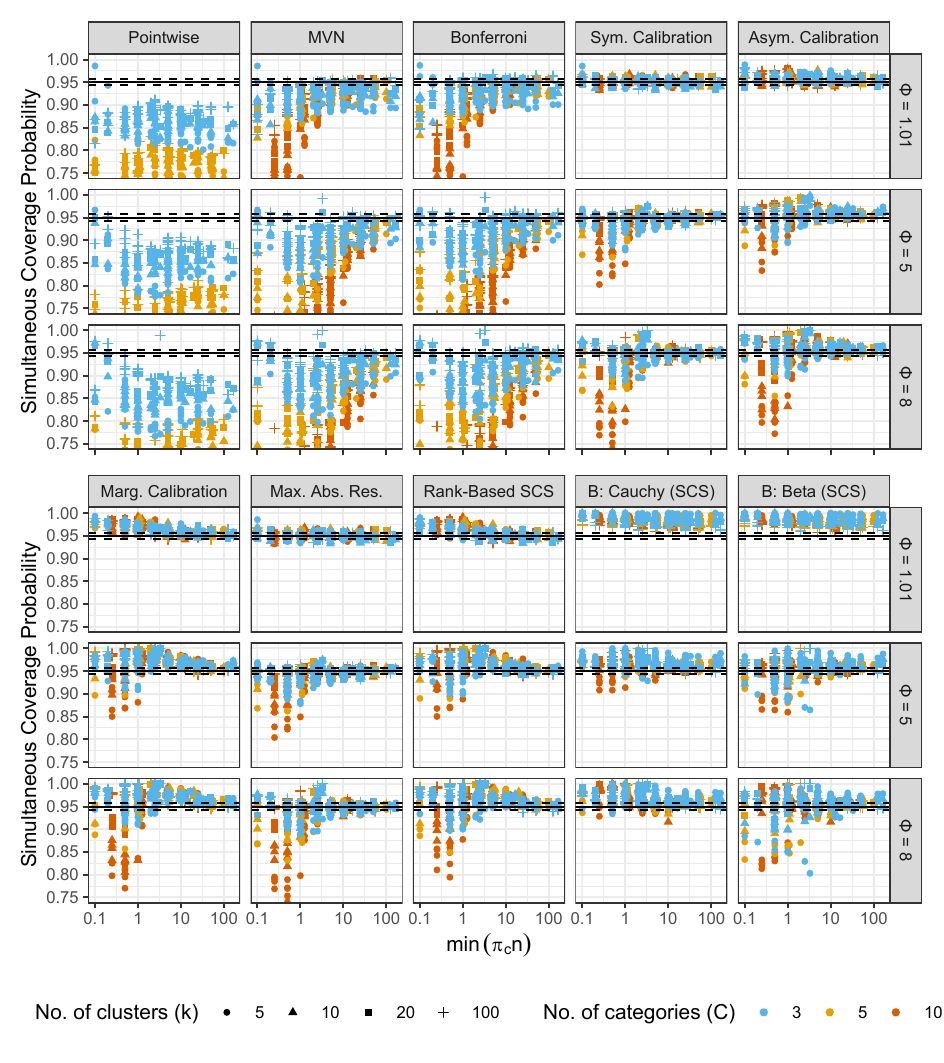}
    \caption{Simulated simultaneous coverage probability of the ten selected methods for all scenarios and settings. The color indicates the number of categories $C$ in the true probability vector and the shape represents number of clusters. The nominal level of 0.95 is represented by the solid horizontal line, and the dashed lines represent the Monte Carlo error of the simulation ($0.95\pm\sqrt{0.95(1-0.95)/1000}$). The horizontal facets indicate the magnitude of overdispersion ($\phi$).}
    \label{fig:covprob_all}
\end{figure}

Figure~\ref{fig:covprob_all} presents the empirical simultaneous coverage probability for ten methods across the full range of simulation settings. Regarding the Bayesian framework, all three interval constructions were evaluated, but only results for the rank-based SCS approach (using both Beta and Cauchy priors which appear in the facets as B: Beta (SCS) and B: Cauchy (SCS)) are presented here because it proved more promising than the marginal-quantile and mean-centered approaches. The x-axis shows the minimum expected count across all categories and historical studies, $\min(\pi_{c}n)$, on a logarithmic scale. The point shape represents number of clusters $k$, and the colors indicate the number of categories $C$ in the true probability vector ($\bm{\pi}_{\text{true}}$). For better visibility and comparability, the $y$-axis is truncated at 0.75. 

The pointwise and MVN approximations, as well as the normal approximation with the Bonferroni adjustment, perform poorly. These methods yield intervals that are consistently too liberal, often falling far below the nominal 95\% coverage level, particularly as the number of categories increases. However, the latter two converge toward the nominal 0.95 level when the minimum expected counts are large. In contrast, the bootstrap-calibrated and Bayesian methods perform substantially better. The Symmetric Calibration and the Maximum Absolute Studentized Residual (MASR) methods, which yield nearly identical results, achieve coverage close to the nominal level for most scenarios but can be slightly liberal when overdispersion is high and expected counts are low. The Asymmetric and Marginal Calibration, along with the Rank-Based SCS methods, perform more consistently, maintaining coverage at or slightly above the nominal level across almost all settings, with the Marginal Calibration being the most conservative of the three. Although the Asymmetric Calibration, Marginal Calibration, MASR, and Rank-Based SCS methods generally performed better than the direct normal-approximation methods (pointwise, Bonferroni-adjusted and MVN), in some extreme settings with $C=10$, only a few historical clusters ($K=5$ or $10$) and small cluster sizes ($n=10$), even these methods can become substantially liberal, with simultaneous coverage probabilities dropping to about 0.75. However, these scenarios are of limited practical relevance, because they correspond to highly sparse multinomial settings in which several categories frequently contain only zero counts. In such situations, the underlying category structure would often be too sparse to support a meaningful analysis. 

The two Bayesian priors also perform well, but are conservative when expected counts are low, no overdispersion is present and especially when the number of historical clusters is small ($K=5$). When overdispersion is introduced, the Cauchy prior tends to be slightly conservative while remaining close to the nominal level, whereas the Beta prior tends to be somewhat liberal at low expected counts. However, with overdispersion present, both Bayesian approaches generally converge to the nominal level as the minimum expected counts increase. 

\begin{figure}[!tbp]
    \centering
    \includegraphics[width=\linewidth,keepaspectratio]{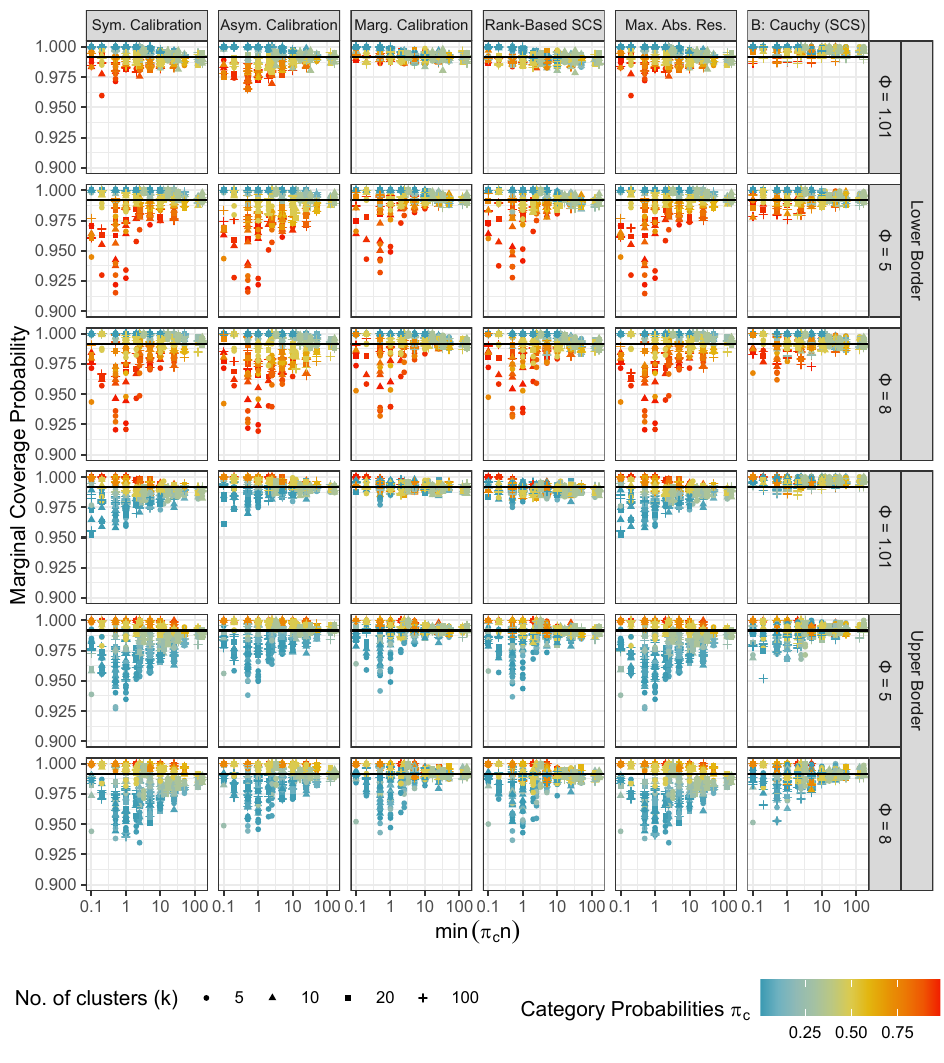}
    \caption{Assessment of equal tail probabilities for the lower and upper historical control limits of the six best-performing methods in three-category ($C=3$) scenarios. The panel groups labeled `Lower Border' and `Upper Border' on the right show $P(y_c \ge L_c)$ and $P(y_c \le U_c)$, respectively. Each point represents one bound-specific probability for one category. The horizontal black line indicates the nominal Bonferroni-adjusted target of $1 - \alpha/(2C) \approx 0.9917$.}
    \label{fig:eqt_lu}
\end{figure}

To further assess whether the intervals achieve approximately equal tail probabilities, Figure~\ref{fig:eqt_lu} examines the marginal probabilities $P(y_c \ge L_c)$ and $P(y_c \le U_c)$ for the six most promising methods in the three-category scenarios ($C=3$), grouped by the right-hand facet labels `Lower Border' and `Upper Border', respectively. If the intervals have approximately equal tail probabilities, both probabilities should be close to the nominal Bonferroni-adjusted target of $1 - \alpha/(2C) \approx 0.9917$, indicated by the horizontal line. A method is considered conservative for a given bound if its probability is above this line and liberal if it is below.

Across most frequentist methods, a systematic pattern emerges. For the lower bound, coverage is often too high (conservative) for categories with low probabilities ($\pi_c$) and too low (liberal) for categories with high probabilities. The inverse is observed for the upper bound, which tends to be liberal for low-$\pi_c$ categories and conservative for high-$\pi_c$ categories. This asymmetry diminishes as the minimum expected counts increase but is exacerbated by higher overdispersion.

The Marginal Calibration and Rank-Based SCS methods perform best in this regard, demonstrating the least variation and the most rapid convergence to the nominal coverage levels. The Symmetric Calibration and Maximum Absolute Studentized Residual methods show the highest variability, confirming their tendency to produce less balanced intervals. The Asymmetric Calibration performs slightly better, though for category probabilities near 0.5, the upper bound tends to be liberal while the lower bound is conservative. 

The Bayesian method with a Cauchy prior displays a slightly less consistent pattern. The conservative behavior observed in the absence of overdispersion is evident here as well, characterized by general conservatism across all category probabilities. When overdispersion is present, the performance more closely resembles that of the frequentist methods, with coverage converging to nominal levels as the minimum expected count increases. Notably, the Bayesian approach is comparatively less liberal at the lower bound for high category probabilities, and less liberal at the upper bound for low category probabilities.

\section{Example: Histopathological Findings}
\label{sec:example}

To illustrate the application of the proposed methods to real historical control data, we analyzed data provided by BASF on alveolar histiocytosis in the lungs of Crl:WI(Han) rats. Only historical control groups from diet studies were included, and female and male datasets were analyzed separately. The dataset comprised 26 historical control groups in total, with 13 female and 13 male groups. The endpoint of interest was the histopathological finding alveolar histiocytosis in the lung, recorded on an ordinal severity scale with the categories `minimal', `slight', `moderate', and `severe'. However, no severe observations occurred in any historical control group, and this category was therefore not considered further in the analysis. Across historical control groups, the total number of affected animals ranged from 15 to 44 in females and from 22 to 38 in males. For each sex-specific historical dataset, we fitted the multinomial model described above and estimated the overdispersion parameter, yielding $\hat{\phi}=1.875$ for females and $\hat{\phi}=1.104$ for males.

\begin{figure}[!tbp]
    \centering
    \includegraphics[width=\linewidth,keepaspectratio]{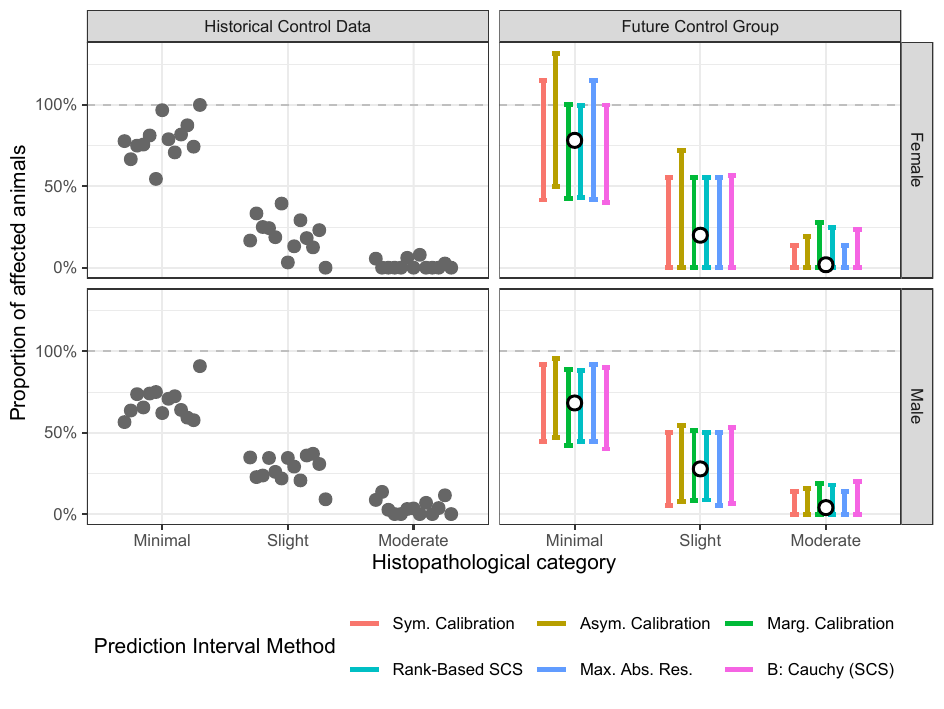}
    \caption{Historical control data and 95\% simultaneous PIs for alveolar histiocytosis in the lungs of Crl:WI(Han) rats, shown separately for males (top row) and females (bottom row). The left panels show the observed proportions in each severity category for the 13 historical control groups per sex. The right panels show the PIs for a future control group of size $m=30$ using the six most promising methods identified in the simulation study. All values are displayed on the proportion scale (as percentages). The black-outlined white circles represent the predicted counts $\hat{y}_c$ expressed as proportions, $\hat{y}_c/m=\hat{\pi}_c$.}
    \label{fig:pat_trial}
\end{figure}

Using these sex-specific historical datasets, we calculated 95\% simultaneous PIs for a future control group of size $m=30$ with all proposed methods. Figure~\ref{fig:pat_trial} displays the intervals on the proportion scale for the six most promising methods identified in the simulation study, alongside the historical control data. The exact interval bounds on the count scale for all methods are reported in Table~\ref{tab:pi_comparison_transposed_no_widths}. As in the simulation study, the asymptotic methods (Pointwise, MVN, Bonferroni) yielded the narrowest intervals for both sexes, whereas the bootstrap-based and Bayesian methods generally produced wider limits, particularly for the less frequent categories. In addition, the Symmetric Calibration and Maximum Absolute Studentized Residual methods produced nearly identical limits, while the Marginal Calibration and Rank-Based SCS methods also gave very similar results.

\begin{table}[!htbp]
\centering
\caption{Predicted number of rats per category ($\hat{y}_c$) and 95\% simultaneous prediction intervals for a future control group ($m=30$), based on sex-specific historical control data on alveolar histiocytosis in the lungs of Crl:WI(Han) rats from diet studies}
\label{tab:pi_comparison_transposed_no_widths}
\begin{tabular}{lccc}
  \toprule
Method & Minimal & Slight & Moderate \\ 
  \multicolumn{4}{c}{\textbf{Female}} \\ \midrule
 \midrule
$\hat{y}_c$ & 23.47 & 5.98 & 0.54 \\ 
   \multicolumn{4}{c}{\textbf{Frequentist Methods}} \\ \midrule
Pointwise & [17.18, 29.77] & [0.00, 12.08] & [0.00, 2.58] \\ 
  MVN & [16.11, 30.83] & [0.00, 13.11] & [0.00, 2.92] \\ 
  Bonferroni & [15.78, 31.16] & [0.00, 13.43] & [0.00, 3.03] \\ 
  Sym. Calibration & [14.23, 32.72] & [0.00, 14.94] & [0.00, 3.53] \\ 
  Asym. Calibration & [15.68, 36.29] & [0.00, 18.40] & [0.00, 4.69] \\ 
  Marg. Calibration & [13.16, 30.27] & [0.00, 15.97] & [0.00, 6.55] \\ 
  Rank-Based SCS & [13.76, 29.98] & [0.00, 15.66] & [0.00, 5.82] \\ 
  Max. Abs. Res. & [14.23, 32.71] & [0.00, 14.93] & [0.00, 3.53] \\ 
   \multicolumn{4}{c}{\textbf{Bayesian Methods}} \\ \midrule
B: Beta (marginal) & [11.00, 30.00] & [0.00, 17.00] & [0.00, 7.00] \\ 
  B: Beta (mean) & [13.00, 33.38] & [0.00, 15.74] & [0.00, 4.55] \\ 
  B: Beta (SCS) & [13.00, 30.00] & [0.00, 16.00] & [0.00, 6.00] \\ 
  B: Cauchy (marginal) & [10.33, 30.00] & [0.00, 18.00] & [0.00, 7.00] \\ 
  B: Cauchy (mean) & [11.93, 34.33] & [0.00, 16.78] & [0.00, 5.00] \\ 
  B: Cauchy (SCS) & [11.00, 30.00] & [0.00, 18.00] & [0.00, 7.00] \\ 
   \midrule
\multicolumn{4}{c}{\textbf{Male}} \\ \midrule
$\hat{y}_c$ & 20.48 & 8.33 & 1.19 \\ 
   \multicolumn{4}{c}{\textbf{Frequentist Methods}} \\ \midrule
Pointwise & [15.01, 25.95] & [3.07, 13.59] & [0.00, 3.48] \\ 
  MVN & [14.05, 26.91] & [2.14, 14.51] & [0.00, 3.89] \\ 
  Bonferroni & [13.80, 27.16] & [1.90, 14.76] & [0.00, 3.99] \\ 
  Sym. Calibration & [13.51, 27.45] & [1.62, 15.03] & [0.00, 4.11] \\ 
  Asym. Calibration & [13.81, 28.46] & [1.91, 16.00] & [0.00, 4.53] \\ 
  Marg. Calibration & [12.29, 27.15] & [2.02, 16.32] & [0.00, 4.67] \\ 
  Rank-Based SCS & [12.78, 27.05] & [2.44, 15.48] & [0.00, 4.53] \\ 
  Max. Abs. Res. & [13.52, 27.45] & [1.63, 15.03] & [0.00, 4.11] \\ 
   \multicolumn{4}{c}{\textbf{Bayesian Methods}} \\ \midrule
B: Beta (marginal) & [12.00, 27.00] & [2.00, 17.00] & [0.00, 6.00] \\ 
  B: Beta (mean) & [12.34, 28.20] & [0.82, 16.00] & [0.00, 4.78] \\ 
  B: Beta (SCS) & [12.00, 27.00] & [2.00, 16.00] & [0.00, 6.00] \\ 
  B: Cauchy (marginal) & [12.00, 27.00] & [2.00, 17.00] & [0.00, 6.00] \\ 
  B: Cauchy (mean) & [12.00, 28.49] & [0.55, 16.28] & [0.00, 4.96] \\ 
  B: Cauchy (SCS) & [12.00, 27.00] & [2.00, 16.00] & [0.00, 6.00] \\ 
   \bottomrule
\end{tabular}
\end{table}

The sex-specific analysis further illustrates the effect of differing overdispersion. Because the estimated dispersion was larger in females than in males, the resulting bootstrap-based intervals were generally wider for the female dataset. This difference was most visible in the upper bounds for the rarer moderate category, where accounting for extra-multinomial variation increased the predicted range of future observations.

The dataset for this example and the code used to compute the prediction intervals are available at \url{https://github.com/sbudig/predint_multnom}.

\section{Discussion}

This article provides a comprehensive description and evaluation of various methods for constructing simultaneous PIs for overdispersed multinomial data, a context for which no formal descriptions currently exist. The primary application is the establishment of historical control limits for endpoints with more than two categories, such as the classification of offspring in DART studies or for histopathological gradings. Our simulation study demonstrates that the choice of method is important, as several simple approaches fail to provide reliable intervals.

The pointwise normal approximation, the Bonferroni adjustment, and the MVN approximation consistently produced intervals that were far too liberal, with empirical coverage probabilities often falling far below the nominal 95\% level. This result is consistent with findings for other types of overdispersed count data. Research on overdispersed binomial and Poisson data has shown that uncalibrated, symmetric Wald-type prediction intervals based on normal approximation can have coverage probabilities below the nominal level, particularly when few historical studies are available. These intervals do not accommodate skewness in the prediction error distribution, which can also lead to unequal lower- and upper-tail error probabilities \cite{Menssen2019, Menssen2024, Menssen2025}.

In contrast, the bootstrap-based and Bayesian methods generally achieved simultaneous coverage probabilities closer to the nominal 95\% level. Among the bootstrap-based approaches, Marginal Calibration and Rank-Based SCS maintained coverage at or slightly above the nominal level across most scenarios, although undercoverage occurred in particularly sparse settings with very small minimum expected counts. The Symmetric and Asymmetric Calibration procedures were adapted to multinomial data from the bootstrap calibration framework implemented in the R package predint \cite{Predint2025}. This framework also provided the basis for the Marginal Calibration method, which calibrates the lower and upper limits separately for each category.

Furthermore, our analysis of the marginal lower- and upper-tail error probabilities revealed important differences in how well the methods achieved equal tail probabilities. The Marginal Calibration and Rank-Based SCS methods performed best again, producing intervals with the most similar lower- and upper-tail probabilities. Other methods, such as the Symmetric Calibration and the Maximum Absolute Studentized Residual, achieved nominal simultaneous coverage but often remained conservative in one tail and liberal in the other for a given category. This shows that achieving nominal overall simultaneous coverage does not necessarily imply equal tail probabilities.

The Bayesian hierarchical models were sensitive to the heterogeneity prior, with conservatism being particularly evident near the multinomial limit, as well as undercoverage in some overdispersed settings. The rank-based intervals under the half-Cauchy prior on $\eta_0$ were closer to the target coverage in settings where the beta prior on the intraclass correlation yielded undercoverage. This empirical comparison does not establish a universally preferable prior. However, the results also demonstrate that, while the Bayesian approach is valuable, careful consideration of hyperpriors is required, which can present challenges for non-statisticians. Further research is needed to identify more suitable prior distributions or hyperparameters for this data situation.

A limitation of our simulation study is the use of the DM distribution for both data generation and the parametric bootstrap. While the DM distribution provides a flexible model for overdispersed multinomial data, the robustness of the proposed methods to alternative data-generating mechanisms remains a topic for future research.

Another fundamental limitation concerns the quality and consistency of the historical data itself. The prediction interval methods proposed here are intended to assess whether the concurrent control group is consistent with the historical control data and therefore it is implicitly assumed that the historical data form a representative reference for the concurrent control group. However, this assumption of a single joint distribution is often questionable. As Zarn et al.~\cite{Zarn2024} note, studies are conducted at different times, by different personnel, and with potential changes in animal strains (genetic drift) or husbandry practices, making it unlikely that all control groups are samples from the identical underlying distribution. This can introduce significant heterogeneity between studies \cite{Dertinger2023}. If heterogeneity between studies is the dominant source of variability, the utility of HCD is greatly diminished. As Menssen and Schaarschmidt~\cite{Menssen2019} discuss, including older, less relevant studies can inflate the estimated overdispersion ($\hat{\phi}$), resulting in PIs that are unnecessarily wide and imprecise. Therefore, in line with the recommendations of Dertinger et al.~\cite{Dertinger2023} and EFSA Panel (PPR) et al.~\cite{EFSA2025}, the methods proposed in this article should only be applied after a thorough evaluation of the quality and stability of the HCD.

For practitioners in toxicology and other fields, our simulation results suggest that Marginal Calibration and Rank-Based SCS are promising approaches for constructing simultaneous prediction intervals for overdispersed multinomial data. Across most settings investigated, both methods achieved simultaneous coverage probabilities close to the nominal level and approximately balanced lower- and upper-tail error probabilities. However, caution is warranted in particularly sparse settings, where undercoverage was observed. These approaches may support the validation of concurrent control groups using historical control data, an application discussed in the recent EFSA scientific opinion \cite{EFSA2025}.

\section*{Acknowledgments}

We thank BASF for providing the historical control data used in the example application.

\clearpage
\bibliographystyle{unsrtnat}
\bibliography{References_PredIntMultDat}

\clearpage
\appendix

\section{Probability vectors used in simulation study}\label{sec:AppMPSS}

The specific probability vectors $\bm{\pi}_{\textnormal{true}}$ used for the data generation process in the simulation study are listed in Tables~\ref{tab:prob_vec_3}, \ref{tab:prob_vec_5}, and \ref{tab:prob_vec_10}. The generator normalizes each input weight vector by its sum.

\begin{table}[!htbp]
\centering
\caption{Probability vectors used for scenarios with $C=3$ categories.}
\label{tab:prob_vec_3}
\begin{tabular}{lccc}
\toprule
Scenario & $\pi_1$ & $\pi_2$ & $\pi_3$ \\
\midrule
1  & 0.33 & 0.33 & 0.33 \\
2  & 0.01 & 0.01 & 0.98 \\
3  & 0.25 & 0.01 & 0.74 \\
4  & 0.49 & 0.02 & 0.49 \\
5  & 0.25 & 0.25 & 0.50 \\
6  & 0.10 & 0.30 & 0.60 \\
7  & 0.02 & 0.03 & 0.95 \\
8  & 0.05 & 0.05 & 0.90 \\
9  & 0.05 & 0.10 & 0.85 \\
10 & 0.05 & 0.15 & 0.80 \\
11 & 0.10 & 0.20 & 0.70 \\
12 & 0.05 & 0.35 & 0.65 \\
\bottomrule
\end{tabular}
\end{table}

\begin{table}[!htbp]
\centering
\caption{Probability vectors used for scenarios with $C=5$ categories.}
\label{tab:prob_vec_5}
\begin{tabular}{lccccc}
\toprule
Scenario & $\pi_1$ & $\pi_2$ & $\pi_3$ & $\pi_4$ & $\pi_5$ \\
\midrule
1  & 0.20 & 0.20 & 0.20 & 0.20 & 0.20 \\
2  & 0.30 & 0.30 & 0.20 & 0.10 & 0.10 \\
3  & 0.44 & 0.22 & 0.11 & 0.11 & 0.11 \\
4  & 0.50 & 0.30 & 0.10 & 0.05 & 0.05 \\
5  & 0.45 & 0.27 & 0.18 & 0.08 & 0.01 \\
6  & 0.70 & 0.10 & 0.10 & 0.05 & 0.05 \\
7  & 0.80 & 0.10 & 0.05 & 0.04 & 0.01 \\
8  & 0.10 & 0.10 & 0.20 & 0.30 & 0.30 \\
9  & 0.11 & 0.11 & 0.11 & 0.22 & 0.44 \\
10 & 0.05 & 0.05 & 0.10 & 0.30 & 0.50 \\
\bottomrule
\end{tabular}
\end{table}

\begin{table}[!htbp]
\centering
\caption{Probability vectors used for scenarios with $C=10$ categories.}
\label{tab:prob_vec_10}
\begin{tabular}{lcccccccccc}
\toprule
Scenario & $\pi_1$ & $\pi_2$ & $\pi_3$ & $\pi_4$ & $\pi_5$ & $\pi_6$ & $\pi_7$ & $\pi_8$ & $\pi_9$ & $\pi_{10}$ \\
\midrule
1  & 0.10 & 0.10 & 0.10 & 0.10 & 0.10 & 0.10 & 0.10 & 0.10 & 0.10 & 0.10 \\
2  & 0.05 & 0.05 & 0.10 & 0.10 & 0.10 & 0.10 & 0.10 & 0.10 & 0.10 & 0.20 \\
3  & 0.05 & 0.05 & 0.05 & 0.05 & 0.10 & 0.10 & 0.10 & 0.10 & 0.10 & 0.30 \\
4  & 0.05 & 0.05 & 0.05 & 0.05 & 0.05 & 0.05 & 0.10 & 0.10 & 0.10 & 0.40 \\
5  & 0.05 & 0.05 & 0.05 & 0.05 & 0.05 & 0.05 & 0.05 & 0.05 & 0.10 & 0.50 \\
6  & 0.025 & 0.025 & 0.025 & 0.025 & 0.05 & 0.05 & 0.05 & 0.05 & 0.10 & 0.60 \\
7  & 0.025 & 0.025 & 0.025 & 0.025 & 0.05 & 0.05 & 0.05 & 0.05 & 0.35 & 0.35 \\
8  & 0.05 & 0.05 & 0.05 & 0.05 & 0.05 & 0.05 & 0.10 & 0.20 & 0.20 & 0.20 \\
9  & 0.025 & 0.025 & 0.025 & 0.025 & 0.05 & 0.05 & 0.20 & 0.20 & 0.20 & 0.20 \\
10 & 0.025 & 0.025 & 0.025 & 0.025 & 0.05 & 0.05 & 0.10 & 0.20 & 0.20 & 0.30 \\
\bottomrule
\end{tabular}
\end{table}


\end{document}